\documentclass[11pt]{article}
\usepackage{hyperref}
\usepackage{times}
\usepackage{geometry}
\usepackage[utf8]{inputenc}
\usepackage{enumitem,amssymb}
\usepackage{graphicx}
\usepackage{fancyhdr}
\usepackage{aas_macros}
\usepackage{mdframed}

\newcommand{\mjup}{$M_{\mathrm{Jup}}$}
\newcommand{\msun}{$M_{\odot}$}

\usepackage[authoryear]{natbib}
\setcitestyle{authoryear,open={(},close={)}}

\mdfdefinestyle{theoremstyle}{
innertopmargin=\topskip,}
\mdtheorem[style=theoremstyle]{lrptextbox}{}

\title{LRP2020: THE OPPORTUNITY OF YOUNG NEARBY ASSOCIATIONS WITH THE ADVENT OF THE \textit{GAIA} MISSION}

\author{Jonathan Gagn\'e, Institut de Recherche sur les Exoplan\` etes, Universit\'e de Montr\'eal\\
Joel Kastner, Rochester Institute of Technology\\
Semyeong Oh, University of Cambridge\\
Jacqueline K. Faherty, American Museum of Natural History\\
John Gizis, University of Delaware\\
Adam Burgasser, University of California, San Diego\\
Evgenya L. Shkolnik, Arizona State University\\
Trevor J. David, Jet Propulsion Laboratory\\
Jinhee Lee, University of Georgia\\
Inseok Song, University of Georgia\\
David Lafreni\` ere, Institut de Recherche sur les Exoplan\` etes, Universit\'e de Montr\'eal\\
Stanimir Metchev, Western University of Canada\\
Ren\'e Doyon, Institut de Recherche sur les Exoplan\` etes, Universit\'e de Montr\'eal\\
Adam Schneider, Arizona State University\\
\'Etienne Artigau, Institut de Recherche sur les Exoplan\` etes, Universit\'e de Montr\'eal}

\begin{document}

\maketitle

\abstract{This white paper proposes leveraging high-quality \textit{Gaia} data available to the worldwide scientific community and complement it with support from Canadian-related facilities to place Canada as a leader in the fields of stellar associations and exoplanet science, and to train Canadian highly qualified personnel through graduate and post-graduate research grants.

\textit{Gaia} has sparked a new era in the study of stellar kinematics by measuring precise distances and proper motions for 1.3 billion stars. These data have already generated more than 1\,700 scientific papers and are guaranteed to remain the source of many more papers for the upcoming decades. More than 900 new age-calibrated young low-mass stars have already been discovered as a direct consequence of the second \textit{Gaia} data release. Some of these may already be host stars to known exoplanet systems or may become so with the progress of the TESS mission that is expected to discover $10^4$ nearby transiting exoplanets in the upcoming decade. This places Canada in a strategic position to leverage \textit{Gaia} data because it has access to several high-resolution spectrometers on 1--4\,m class telescopes (e.g. The ESPaDOnS, SPIRou and NIRPS), that would allow to quickly characterize this large number of low-mass stars and their exoplanet systems. This white paper describes the opportunity in such scientific projects that could place Canada as a leader in the fields of stellar associations and exoplanets.}

\section{Young Stellar Associations are Powerful Age-Calibrated Laboratories}

Young stellar associations are the sparse remnants of individual star formation events, where typically hundreds of stars and brown dwarfs formed together in a molecular cloud of relatively uniform composition, within a short period of time. These collections of stars are thus born on Galactic orbits similar to their parent molecular cloud, and gradually get dispersed by gravitational interactions in the span of a few hundred million years. As such, relatively young ($\lesssim$\,200\,Myr) groups of stars can still be identified from their common 3-dimensional space velocities and their various features caused by young age, even though they are too sparse to be gravitationally bound together.

\begin{figure}
	\centering
	\includegraphics[width=0.965\textwidth]{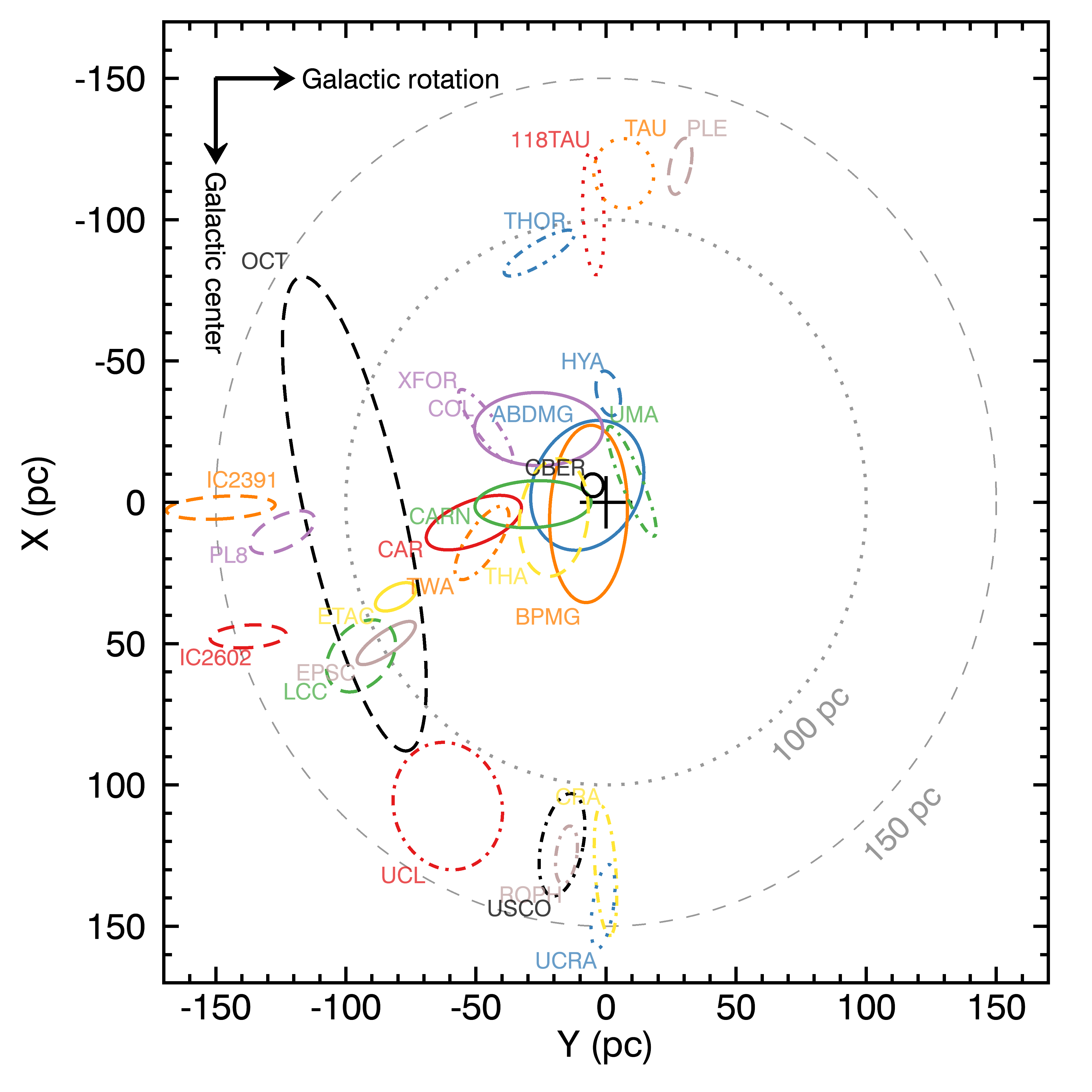}
	\caption{Known stellar associations in the Solar neighborhood. The distribution of each association in Galactic $XY$ coordinates is represented with an ellipsoid model, with the Sun centered (black cross) at ($0$,$0$). The members of the nearest associations are harder to identify because they are spread over the sky, but they can be studied in much more detail due to their proximity.}
	\label{fig:xy}
\end{figure}

\begin{figure}
	\centering
	\includegraphics[width=0.965\textwidth]{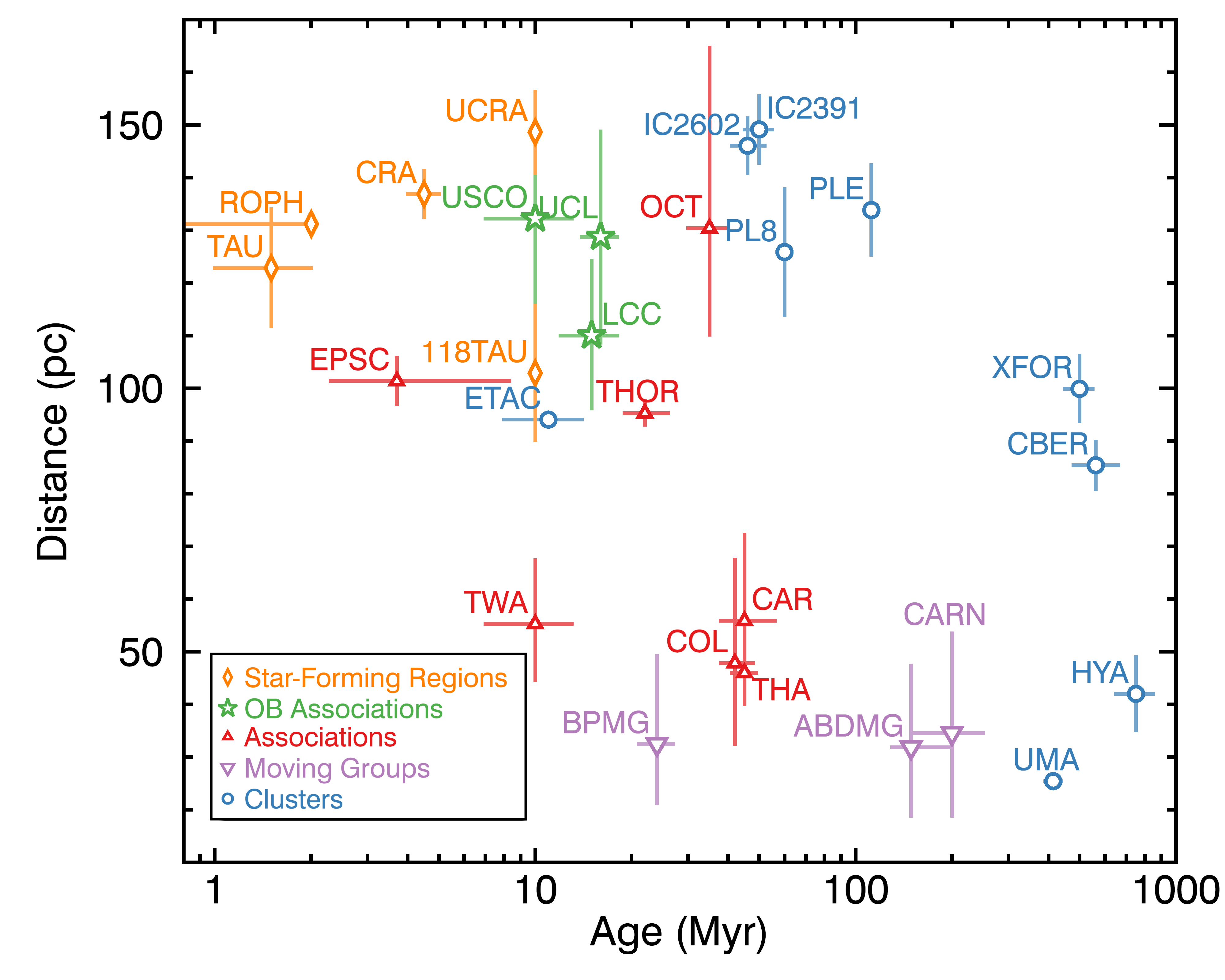}
	\caption{Distances and ages of known stellar associations in the Solar neighborhood. Known associations provide snapshots in time at various different ages, which is crucial to understand how stellar and exoplanet characteristics evolve over time. Discovering new stellar associations of different ages would provide additional age-calibrated stars, and help to map out stellar evolution.}
	\label{fig:ages}
\end{figure}

These similar birth conditions make young stellar associations valuable laboratories to understand the evolution of stars, mainly because they are coeval and of similar chemical composition. Determining the age of a single star with precision is an incredible challenge, because most age-dating methods are imprecise ($>$\,50\%; e.g. \citealt{2008ApJ...687.1264M}) and can only be estimated for certain types of stars. Having access to a large collection of stars with the same age but different masses and temperatures allows us to combine several different age-dating methods and obtain a much more precise age for the association, typically with a 10--20\% precision \citep{2015MNRAS.454..593B}. Currently, a few dozen such associations younger than 1\,Gyr are known within 150\,pc of the Sun (see Figures~\ref{fig:xy} and \ref{fig:ages}).

The precise determinations of stellar ages from membership in a young association make it possible to pose empirical constraints on various stellar properties and better understand their evolution. Some examples include, but are not limited to, the lifetime of accretion disks; the structure of exoplanet system orbits; stellar activity, magnetic fields, UV and X-ray emission; the evolution of stellar angular momentum; the lithium depletion rate; and the cooling rates and cloud properties of brown dwarfs and gas giant exoplanets.

Young stellar associations also present the opportunity to study how the initial mass function, particularly its low-mass end, depends on the stellar formation environment. The initial mass function is one of the most fundamental measurements to constrain the physics of star formation. It is measured by compiling the distribution of masses for a collection of astrophysical objects born from the same molecular cloud, at the time of birth. So far, studies have shown that stellar associations farther than $\sim$\,150\,pc from the Sun have similar initial mass functions above 0.5\,\msun\ \citep{2008ApJ...683L.183A}. The low-mass end of this distribution is still poorly constrained \citep{2012ApJ...753..156K,2014AJ....147..146K,2017ApJS..228...18G,2019ApJS..240...19K} because lower-mass members are fainter. As a consequence, it is still unclear what is the lowest-mass object that can be formed by the stellar formation mechanism (models estimate a minimum mass of $\sim$\,1--10\,\mjup; \citealp{1976MNRAS.176..367L,2004AA...427..299W}), and whether a fraction of the lowest-mass substellar objects ($\lesssim$\,12--13\,\mjup) may have formed by a different process such as the ejection of gas giant exoplanets in a multiple star system. The nearest young stellar associations have the potential to further our understanding of the low-mass end of the initial mass function, because their lowest-mass members are detectable with all-sky near-infrared surveys have masses well below the deuterium burning limit ($\sim$\,13\,\mjup), and into a mass regime that overlaps with giant planets \citep{2013ApJ...777L..20L,2015ApJ...808L..20G,2016ApJ...822L...1S,2016ApJ...821L..15K,2017ApJ...841L...1G,2018ApJ...854L..27G}. Improving the current census of nearby associations and mapping out their spatial extensions with \textit{Gaia} has the potential to benefit searches for such planetary-mass objects.

\section{The Challenges in Studying Young Associations}

\begin{figure}
	\centering
	\includegraphics[width=0.965\textwidth]{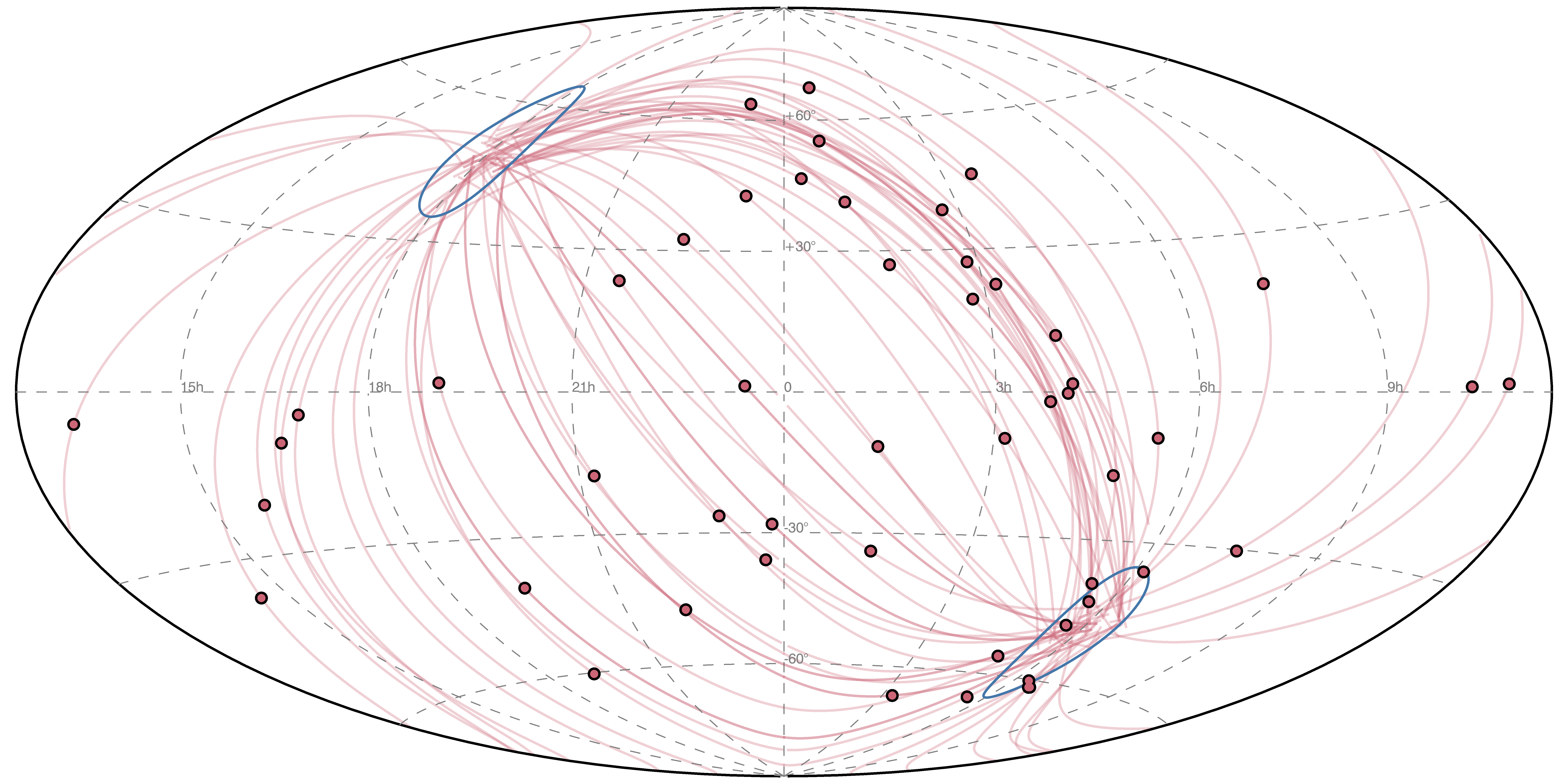}
	\caption{Sky distribution of known AB~Doradus members. AB~Doradus ($\approx$\,150\,Myr) is one of the nearest known young stellar associations, and as a consequence its members are spread over the sky. This makes the identification of its members challenging because 6-dimensional kinematics must be measured before membership can be assessed with confidence. Pink lines represent the past and future trajectories of each member and the blue lines represent their apparent point of origin and destination, a consequence of the common 6-dimensional motion of the young association relative to that of the Sun.}
	\label{fig:abdmg_pm}
\end{figure}

The members of nearby and sparse stellar associations are challenging to identify because they can be spread across a significant fraction of the sky (e.g., Figure~\ref{fig:abdmg_pm}). However, they are also valuable laboratories because we can study each member in much greater detail due to their proximity. As their members’ sky positions alone contain no useful information about their membership, those must be combined with precise 3-dimensional kinematics before their membership becomes clear. Such 3-dimensional velocities require measuring proper motions (the motion of a star projected on the sky), radial velocities from spectroscopy and distances from trigonometric parallaxes. Obtaining these measurements for a large number of stars is challenging because of the significant resources that they require; proper motions and parallaxes require precise astrometric monitoring for a few years, and radial velocity measurements require high-resolution spectroscopy. Furthermore, multi-object spectroscopy is typically impossible for the nearby associations because of the large sky area over which their members are spread.

Statistical methods have been developed to identify potential members of nearby young associations before their 6-dimensional kinematics are fully measured, in order to narrow the lists of objects that need to be followed up to assess their membership with more confidence (e.g., \citealp{2005ApJ...634.1385M,2013ApJ...762...88M,2014ApJ...783..121G,2017AJ....153...95R,2018ApJ...856...23G}). All of these methods suffer from non-negligible rates of false positives (non-members that appear to have the right kinematics) and false negatives (true members that are overlooked). These statistical misses are most important for the nearest young associations such as AB~Doradus and $\beta$\,~Pictoris \citep{2004ARAA..42..685Z}, which are also the most promising because once their members are found, we can study them in greater detail from their proximity. The lowest-mass members of these associations ($<$\,0.3\msun), which would constitute $\sim$\,50\% of their population based on the initial mass function of the Galactic neighborhood, were largely missing until very recently because few kinematic data were available for a large number of such low-mass stars. This prevented a characterization of the low-mass tail of their initial mass function (e.g., \citealp{2014AJ....147..146K,2017ApJS..228...18G,2017AJ....154...69S}).

\section{The \textit{Gaia} Mission Unlocked the Doors to Low-Risk, High-Reward Projects}

\begin{figure}
	\centering
	\includegraphics[width=0.865\textwidth]{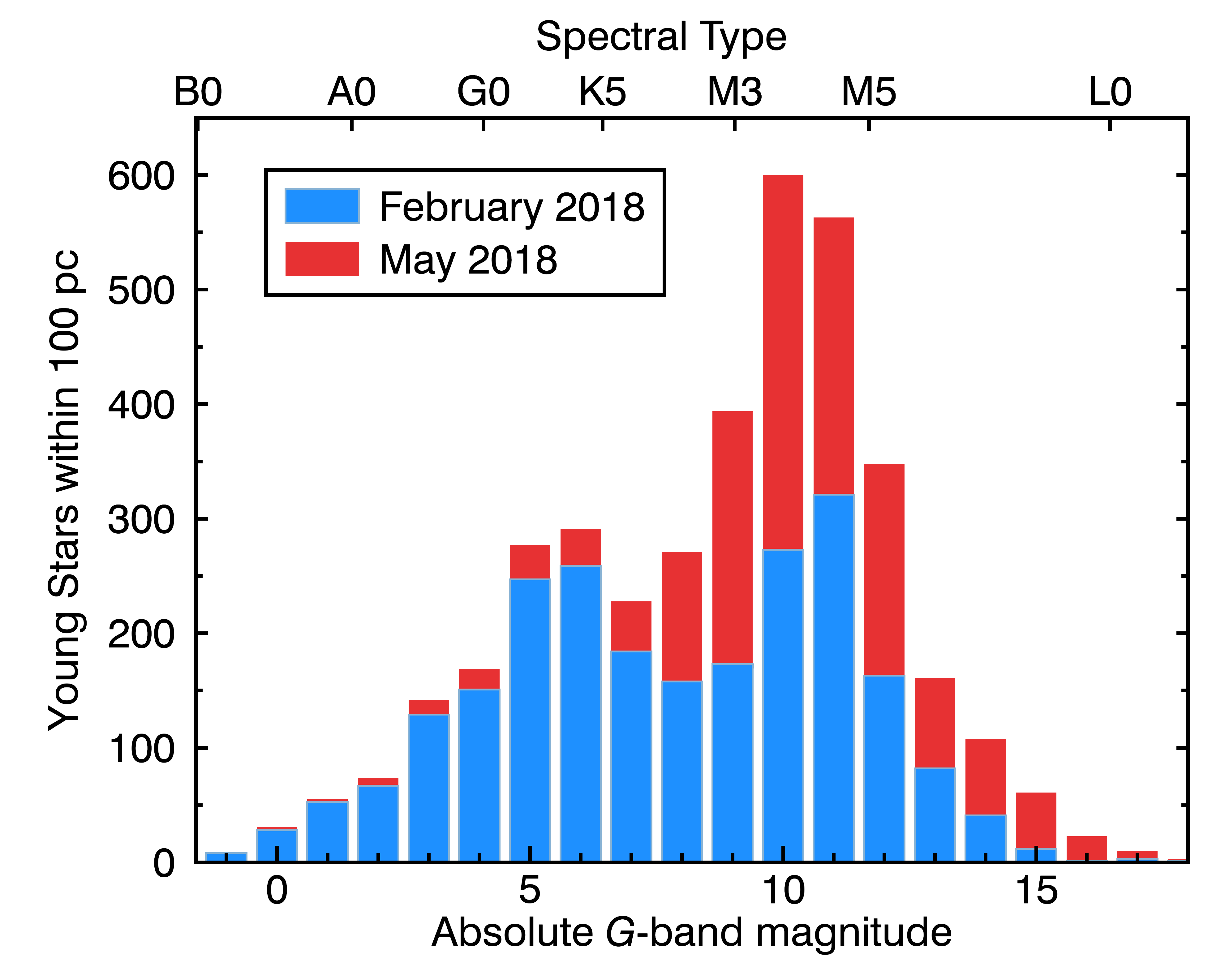}
	\caption{\textit{Gaia} visible $G$-band magnitude distribution of known stellar association members within 100\,pc of the Sun before the second \textit{Gaia} data release (blue), compared to the same distribution one month after the second data release (red). Kinematic measurements from \textit{Gaia} allowed us to quickly boost the number of age-calibrated young low-mass stars by a factor two; Canada is in a significantly strategic position to characterize these newly discovered young low-mass stars with facilities such as ESPaDOnS on the CFHT and SPIRou.}
	\label{fig:gaia_effect}
\end{figure}

In April of 2018, The \textit{Gaia} mission \citep{2016AA...595A...1G} published its second data release \citep{GaiaCollaboration:2018io} which included trigonometric distance measurements for 1.3 billion stars and radial velocity measurements for 7.2 million brighter stars. This represents a 10\,000-fold improvement in numbers of sources, and a 300-fold improvement in precision over the previous census of trigonometric distances from the Hipparcos mission \citep{1997AA...323L..49P}, which had fueled the study of stellar kinematics for decades. The \textit{Gaia} spacecraft is still gathering measurements that will be released in future incremental data releases. This represents a major overhaul in terms of the quality and quantity of kinematic data readily available to the wide scientific community, in the Solar neighborhood and beyond. In the month following the second \textit{Gaia} data release, 900 new low-mass stars that are likely members of young associations were identified (\citealt{2018ApJ...862..138G}; see Figure~\ref{fig:gaia_effect}), effectively doubling the number of known nearby, age-calibrated low-mass stars.

New measurements from \textit{Gaia} already allowed the community to identify several new groups of stars \citep{2017AJ....153..257O,2018arXiv180409058F,2018ApJ...865..136G} that were too sparse to have been identified before. These searches for new associations still remain incomplete, and we should expect that several more have yet to be found. Spatial extensions of known associations have also been identified with \textit{Gaia} data, including tidal tails for the well-studied Hyades, Coma Berenices and Praesepe nearby clusters \citep{2019AA...621L...2R,2019ApJ...877...12T,Roser:2019hb}.

In parallel to these scientific developments, the exoplanet community is currently discovering new exoplanets at a significant rate (over 4\,000 exoplanets are confirmed to date\footnote{See \url{https://exoplanetarchive.ipac.caltech.edu}}). Some of these exoplanet systems are members of young stellar associations with known ages (e.g., \citealp{2014ApJ...787....5N,2015ApJ...806..254A,2017AJ....153...64M}), but most exoplanet systems have not been investigated for young association membership. Other techniques have been used to investigate the ages of exoplanet populations (e.g., \citealp{2005AA...443..609S,2015MNRAS.452.2127S}), but leveraging the known ages of young associations could potentially provide a large number of additional age-calibrated exoplanet systems. This overlap between the fields of exoplanets and young associations will only become more significant with the ongoing TESS mission \citep{2014SPIE.9143E..20R}, which is expected to discover $10^4$ new nearby transiting exoplanets \citep{2018arXiv180711129H}.

These considerations place the astrophysics community in a regime where the amount of high-quality data can fuel more than a decade of low-risk, high-reward projects, making this a strategic time to train new experts in the fields of young stars and exoplanets.

\section{Canada is Strategically Positioned to Lead the Study of Young Associations}

As a consequence of the scenario described above, the Canadian astrophysics community could play a forefront role in furthering the study of young associations in a way that is synergic with the fields of stellar astrophysics and exoplanets. This could be implemented with the following strategies:

(1) Allocating funds to the training of graduate and post-graduate researchers for the analysis of available \textit{Gaia} and \textit{TESS} data, in the form of grants or research chairs.

(2) Maintaining Canadian access to mid- to high-resolution optical and infrared spectrometers mounted on 1--4\,meter telescopes in both the Northern and Southern hemispheres to allow a detailed characterization of newly discovered low-mass members of young associations. This includes instruments such as VROOM at the Observatoire du Mont-Mégantic, ESPaDOnS at the Canada-France-Hawaii telescope (CFHT); GNIRS and GMOS-N at the Gemini-North telescope; and GMOS-S and Flamingos-2 at the Gemini-South telescope.

(3) Maintaining Canadian involvement in instruments such as the SPectropolarimètre InfraRouge (SPIRou; \citealt{2018haex.bookE.107D}) at the CFHT and the Near-Infrared Planet Searcher (NIRPS; \citealt{2017Msngr.169...21B}) at the ESO~3.6\,m telescope to keep Canada as a leader in the discovery and characterization of exoplanets in the Solar neighborhood.

These particular resources will allow Canada to produce forefront scientific papers that will have a profound impact on our understanding of how stars and their exoplanet systems evolve with age, with low-cost facilities that are mostly already in place. Mid- to high-resolution optical spectroscopy on 1--4\,m telescopes will be crucial to study the chemical composition of the hundreds of newly discovered young low-mass stars (e.g., Li, Na, H$\alpha$), and their rotational velocities to understand how these quantities evolve in time and in different birth environments. This will also be useful to measure radial velocities for stars too faint to have a \textit{Gaia} radial velocity measurement, therefore completing their 6-dimensional kinematics.

There is currently orders of magnitude more data that can be analyzed by the scientific community in a few years, and the Canadian community includes world experts in the fields of young stars, stellar associations, substellar objects and exoplanets. It would be strategic to take this opportunity and train more students and researchers while leveraging the available \textit{Gaia} data to lead scientific projects that have an unusually low risk of null results: the \textit{Gaia} mission has already published a significant amount of data which met or exceeded its design goals \citep{GaiaCollaboration:2018io}. This very high quality data will allow researchers to identify the members of young associations with a rate of false positives that is orders of magnitude smaller than before \citep{2018ApJ...862..138G}.

SPIRou is the first instrument combining precision radial velocity (pRV; 1\,ms$^{-1}$ accuracy) through high-resolution spectroscopy ($R\sim70\,000$ over the 0.98--2.45\,$\mu$m domain) with polarimetric capabilities in the near-infrared. The combination of polarimetry and pRV is key in disentangling activity-induced jitter of the host star from planetary signals. SPIRou has seen first light at the CFHT in April 2018, and is currently being used for the SPIRou Legacy Survey (SLS). The 300\,nights allocation of the SLS will be used to probe the architecture of M-dwarf planetary systems as well as magnetic field topologies in embedded protostars.

NIRPS is a pRV instrument operating over the 0.98--1.80\,$\mu$m range that will be used in parallel with the existing HARPS optical pRV instrument \citep{2002Msngr.110....9P}. The simultaneous use of optical and infrared pRV is another approach for disentangling activity and planetary signals. NIRPS is slated for first light in mid-2020 at the La Silla telescope. While hosted at an ESO facility, Canadian astronomers have gained access to a very large amount of time on NIRPS through guaranteed time observation (GTO). The NIRPS consortium has been awarded a GTO allocation corresponding to 40\% of the La Silla 3.6\,m telescope over 5 years; this amount of time is  larger than all the time awarded to Canadian astronomers at CFHT over the same period of time. As Canadian astronomers from contribution NIRPS institutions (McGill, Bishop's, RMC Kingston, UdeM, UBC) make nearly half of the NIRPS science core science team, this amounts to the largest time allocation on any 4\,m class telescope for Canadians.

While SPIRou and NIRPS were largely developed for planet searches, they are the best type of instrument for physical characterisation and guiding stellar models in the low-mass stellar and substellar regime. This is especially true of young stars because they are more active \citep{2008ApJ...687.1264M} and the combination of infrared and optical pRVs is a powerful means to distinguish planetary signals from stellar activity (e.g., \citealt{2016MNRAS.459.3565V}). Their very high resolution resolves molecular and atomic line profiles entirely (i.e., a larger resolution would not provide more information), and their broad wavelength coverage spans the peak of the spectral energy distribution of low-mass stars and brown dwarfs. The observation of such low-mass astrophysical objects at this resolution is still in its infancy, and Canadian astronomers are well poised to take advantage of these facilities.

\pagebreak

\begin{lrptextbox}[How does the proposed initiative result in fundamental or transformational advances in our understanding of the Universe?]

The scientific projects described in this white paper have the opportunity to drastically improve our ability to determine the ages of stars and their planetary systems through the study of coeval stellar associations. Understanding how the properties of stars and exoplanet systems evolve with time will provide immediate benefits in (1) targeted searches for exoplanets, (2) characterizing known exoplanet systems, (3) improving our understanding of stellar evolution as described in this white paper, and (4) better understand the formation history of the Solar neighborhood and the wider Galactic disk.

\end{lrptextbox}

\begin{lrptextbox}[What are the main scientific risks and how will they be mitigated?]

The main scientific risks are associated with null scientific results because this white paper does not require new facilities or instruments. As outlined in the main text above, the risk of null scientific results is low given the abundance and high quality of publicly available \textit{Gaia} data that is ripe for analysis.

\end{lrptextbox}

\begin{lrptextbox}[Is there the expectation of and capacity for Canadian scientific, technical or strategic leadership?] 

There is capacity for Canadian scientific strategic leadership as the Canadian astrophysics community includes world leaders in the fields of young stars and exoplanets. Canada can therefore play a forefront role in developing knowledge on the topic of young stellar associations, and immediately apply these innovations to the fields of exoplanets and stellar evolution. Canada has access to facilities such as the CFHT telescope and the Gemini-North and South telescopes, and plays an active role in international missions such as SPIRou that will allow an observational follow-up synergic with \textit{Gaia} data to both further our understanding of young stellar associations and leverage the resulting improved age calibrations for the study of exoplanets.

\end{lrptextbox}

\begin{lrptextbox}[Is there support from, involvement from, and coordination within the relevant Canadian community and more broadly?] 

There is current support from the Canadian community in the form of guaranteed Canadian access to the CFHT, Gemini-North and Gemini-South telescopes, SPIRou and NIRPS. There is Canadian and worldwide support in the form of national and international scientific collaborations with experts on the topics of young stars and exoplanets.

\end{lrptextbox}

\begin{lrptextbox}[Will this program position Canadian astronomy for future opportunities and returns in 2020-2030 or beyond 2030?]

This program will position Canada as one of the leaders in determining the ages of stars and exoplanets, which importance has become central to understand the population properties of stars and characterize the radius and thermal emission evolution of giant exoplanets. It will place Canadian researchers in a strategic position to lead further scientific projects based on \textit{Gaia} data that will be a significant legacy for the upcoming decades; Since its release in April 2018, the second \textit{Gaia} data release has already been cited more than 1\,700 times in peer-reviewed papers. Some of these future potential projects include the discovery and characterization of exoplanets with the astrometric method and the direct determination of stellar, substellar and exoplanet masses from the detection of projected acceleration.

\end{lrptextbox}

\begin{lrptextbox}[In what ways is the cost-benefit ratio, including existing investments and future operating costs, favourable?] 

The cost-benefit ratio is especially favorable for this scientific topic because (1) \textit{Gaia} data is already accessible and will continue to be accessible to the wide astrophysics community; (2) Canada already has access to facilities and is involved with ongoing projects (e.g., ESPaDOnS, SPIRou and NIRPS) that are synergic with the proposed science; (3) the cost of financing graduate and post-graduate researchers and maintaining facilities to develop this field is low and it will significantly benefit Canada by placing it as a leader in the fields of stellar associations and exoplanets, as well as allowing it to train a generation of highly skilled researchers.

\end{lrptextbox}

\begin{lrptextbox}[What are the main programmatic risks
and how will they be mitigated?] 

The programmatic risks are very low for the scientific topics described in this white paper as they (1) rely on already existing facilities; (2) are associated with very low risks of null scientific results; and (3) will result in the training of scientific researchers independently of the scientific results.

\end{lrptextbox}

\begin{lrptextbox}[Does the proposed initiative offer specific tangible benefits to Canadians, including but not limited to interdisciplinary research, industry opportunities, HQP training, EDI, outreach or education?] 

This white paper bears direct relevance to the training of highly qualified personnel as it will allow graduate and postgraduate researchers to analyze a large amount of high-quality data with various numerical and data learning methods, therefore equipping them with solid scientific methodology and highly transferable skills. The generation of knowledge on the formation history of the Solar neighborhood as well as the characterization of exoplanets are especially useful for outreach activities because the wider public is generally very receptive to these scientific topics.

\end{lrptextbox}

\end{document}